# Answering Table Queries on the Web using Column Keywords

Rakesh Pimplikar[*]
IBM Research
New Delhi, India
rakesh.pimplikar@in.ibm.com

Sunita Sarawagi
IIT Bombay
Mumbai, India
sunita@iitb.ac.in

## ABSTRACT

We present the design of a structured search engine which returns a multi-column table in response to a query consisting of keywords describing each of its columns. We answer such queries by exploiting the millions of tables on the Web because these are much richer sources of structured knowledge than free-format text. However, a corpus of tables harvested from arbitrary HTML web pages presents huge challenges of diversity and redundancy not seen in centrally edited knowledge bases. We concentrate on one concrete task in this paper. Given a set of Web tables $T_1, \ldots, T_n$, and a query $Q$ with $q$ sets of keywords $Q_1, \ldots, Q_q$, decide for each $T_i$ if it is relevant to $Q$ and if so, identify the mapping between the columns of $T_i$ and query columns. We represent this task as a graphical model that jointly maps all tables by incorporating diverse sources of clues spanning matches in different parts of the table, corpus-wide co-occurrence statistics, and content overlap across table columns. We define a novel query segmentation model for matching keywords to table columns, and a robust mechanism of exploiting content overlap across table columns. We design efficient inference algorithms based on bipartite matching and constrained graph cuts to solve the joint labeling task. Experiments on a workload of 59 queries over a 25 million web table corpus shows significant boost in accuracy over baseline IR methods.

## 1. INTRODUCTION

We consider the following structured Web search problem. A user wants to assemble a multi column table representing either entities with an optional set of attributes, or multi-ary relationships among entities. He expresses his query as sets of keywords, one set for each column he wishes to see in the answer table. Examples include, single column keyword queries like "Mountains in North America" to retrieve names of entities; two column keyword queries like "Pain killer | Side effects" to retrieve instances of relationship between two entities; three column keyword queries like "Cheese name | Country of origin | Milk source" to find entities along with values of two attributes. We tap the large number of organically created tables on the Web to answer such queries. In a recent 500 million pages Web crawl, we conservatively estimated that over 25 million tables express structured information. Similar statistics have been reported elsewhere on other web crawls [4, 3, 7]. Each table contributes valuable facts about entities, their types, and relationships between them, and does so in a manner that is considerably less diverse and less noisy, compared to how facts are expressed in free-format text on the general Web. This makes web tables an extremely valuable resource for answering structured queries on the Web.

The power of Web tables to answer ad hoc structured queries has largely been untapped. One exception is the Octopus system [2] that supports keyword queries such as "Mountains in North America" and returns a ranked list of tables that match those keywords and thereafter depends on multi-round user interactions to assemble an answer table. The system is targeted for topic queries, and is not suitable for *ad hoc* retrieval of relationships or specific attribute sets of an entity type. Multi-column keyword queries provide a unified mechanism to query for entities, attributes of entities, and relationships, while being only a small extension of the familiar search box keyword queries.

We present the design of an end to end system called WWT that takes as input a large corpus $D$ of Web tables and in response to a query $Q$ with q column keyword sets $Q_1, \ldots, Q_q$ returns as answer a single q column table. Unlike standard IR systems, our goal is not just to present a ranked list of matching Web tables from $D$, but to extract relevant columns from several matching web tables and consolidate them into a single structured table. An illustrative scenario is shown in Figure 1. The user is interested in compiling a table about explorers listing their names, nationality and areas explored. He submits three sets of column keywords $Q_1, Q_2, Q_3$ as a query as shown in the top left part of the figure. We show snippets of three of the several tables that match the query keywords. With each table snippet we also show some text, that we call *context* that was extracted from around the table in the web document that contained the table. WWT processes these tables to get a final consolidated table as shown as a snippet in the top right corner of the figure.

A crucial challenge in converting the web tables into a consolidated answer table is deciding if a web table is rel-

---

[*]Most of the work was done when the author was a student at IIT Bombay.

## Figure 1

**Column Descriptors**

| Name of Explorers | Nationality | Areas Explored |
|---|---|---|
| $Q_1$ | $Q_2$ | $Q_3$ |

**Answer Table**

| Name of Explorers | Nationality | Areas Explored |
|---|---|---|
| Vasco da Gama | Portuguese | Sea route to India |
| Abel Tasman | Dutch | Oceania |
| Christopher Columbus | | Caribbean |
| ... | ... | ... |

**Web Table 1**

List of explorers - Wikipedia, the free encyclopedia

| Name | Nationality | Main areas explored |
|---|---|---|
| Abel Tasman | Dutch | Oceania |
| Vasco da Gama | Portuguese | Sea route to India |
| Alexander Mackenzie | British | Canada |
| ... | ... | ... |

**Web Table 2**

This article lists the explorations in history. For the documentary 'Explorations, powered by Duracell', see Explorations (TV)

| Exploration (Chronological order) | Who (explorer) |
|---|---|
| Sea route to India | Vasco da Gama |
| Caribbean | Christopher Columbus |
| Oceania | Abel Tasman |
| ... | ... |

**Web Table 3**

Other Formal Reserves 1.3 Forest Reserves under the Forestry Act 1920

All areas will be available for mineral exploration and mining evant, and if so, matching the columns of the table to the query columns. We call this the Column mapping task. In Figure 1, the task should label Table 1 as relevant and map its columns consecutively to $Q_1, Q_2, Q_3$, label Table 2 as relevant and map column 1 to $Q_3$ and column 2 to $Q_1$, and label Table 3 as irrelevant. The Column mapping task is more challenging than relevance ranking entire tables because the amount of column specific information from both the query and tables is limited. From the query side, the only information is a set of keywords, and from the web table side the only obvious column-specific information is in the header of the table. Headers of tables on the Web are inadequate in several ways: most Web tables do not use the designated HTML tag for indicating headers (80% in our corpus) and are identified via imperfect automated methods, many tables have no headers (18% in our corpus), header text is often uninformative (e.g. "Name"), many tables have multiple headers but without any clearcut semantics (do they represent split phrases as in Table 1 of Figure 1, or additional information as in Table 2, or a title as in Table 3?). The context text of a table is useful but it does not give column-specific information. How can matches in the context, common to all columns of a table, help us discriminate one column from another? We make a number of contributions to address these challenges.

| Forest reserves | | |
|---|---|---|
| ID | Name | Area |
| 7 | Shakespeare Hills | 2236 |
| 9 | Plains Creek | 880 |
| 13 | Welcome Swamp | 168 |
| ... | ... | ... |

Figure 1: An example column description query.

*Contributions.* We propose a novel method to combine the clues in one column's header with the signals from the rest of the table (such as its context, and body) through a two-part query segmentation model. We show that the segmentation model is more precise than the IR practice of weighted sum of whole string matches along each field. Second, we show how to exploit the content overlap of a column with columns of other tables to help tables with poorly matched headers. We use an elegant formulation based on graphical models to jointly map all columns of all tables to the query table. We discuss a number of non-trivial choices in the design of the graphical model, including the choice of the variables, the node potentials that incorporate diverse clues from the table and the corpus, and, edge potentials that exploit content overlap while being robust to the presence of several irrelevant web tables. Finally, we design efficient algorithms to solve the joint column mapping task. Experiments on a workload of 59 queries show a boost of F1 accuracy from 65% using baseline IR method to 70% using our approach.

*Outline.* We present an end to end description of WWT in Section 2. In Section 3 we describe the column mapper. In Section 4 we present algorithms to find the highest scoring mapping in the graphical model. Experiments, related work, and conclusions appear in Sections 5, 6 and 7 respectively.

## 2. ARCHITECTURE

Figure 2 shows the architecture of WWT. We first describe the main functions performed offline and then the process of answering a query online.

### 2.1 Offline Processing

We start with a Web Crawl from which we extract all data within the HTML table tags. However, not all of these are data tables because the table tag is often used for layout and for storing artifacts like forms, calendars, and lists. We depend on heuristics to retain tables that present relational information in a vertical format. On a 500 million pages Web crawl, we get an yield of 25 million data tables, which is roughly 10% of the table tags in the crawl. Similar statistics have been reported in [4, 3, 7]. A machine learning based approach [7] could have been more accurate but we did not have enough labeled data at the time. Instead, we decided to rely on query time relevance judgments to filter away non-data tables. Each extracted table is then processed and associated with the following two types of information:

#### 2.1.1 Headers

Even though the HTML language has a designated tag ("th") for headers, only 20% of the 25 million Web tables used it. Most others depended on visual, font, and content markers to indicate headers. After exploring several Web tables in an active learning framework, we designed an algorithm that marks headers as follows. The rows of a table are assumed to consist of zero or more title rows (for example Table 3 of Figure 1 has one title row), followed by zero or more header rows, followed by body rows. We scan rows sequentially from the top as long as we find rows different from most of the rows below it in terms of formatting (use of bold face, italics, underline, capitalization, code, and header tags), layout (background color, CSS classes), or content (textual header with numeric body, number of characters). A 'different' row is labeled a title if all but the first column is non-empty. Else, we label that row as a header. All subsequent rows are labeled header as long as they are similar to the first header row and different from the rows below them. We stop as soon as a row fails this test. Over two thousand Web tables, we found that this heuristic failed only in five cases. In our corpus as per this method, 60% tables had one header row, 18% had no header, 17% had two header rows and the remaining 5% had more than two header rows. We could have used a less heuristic approach, say based on CRFs [14] but we found this approach to be very fast, and sufficiently accurate.

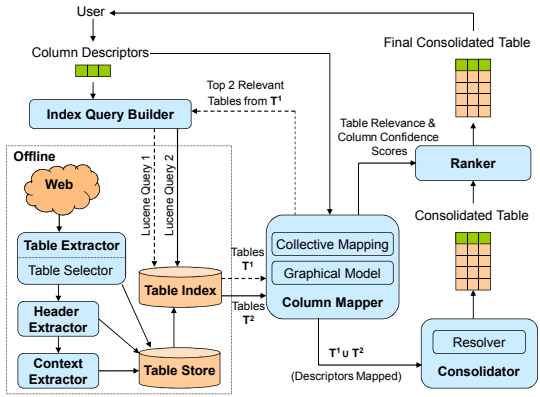

**Figure 2: System architecture.**

### 2.1.2 Context

The context is meant to include all the text in the parent document that tells us what the table is about. But this notion is much less objective than a header. So, we followed a policy of being generous in including text snippets in the context and, attaching a score to each text snippet for use during query processing. We depend on the DOM tree $d$ of the parent document for extracting candidate snippets. Let $T$ be the node in $d$ containing the table. Any text node $x$ that is a sibling of a node on the path from $T$ to the root of $d$ is added to the context. The score attached with $x$ is calculated from two types of factors (1) the edge distance in tree $d$ between $x$ and $T$ and whether $x$ is a right or left sibling of the path nodes, and (2) the relative frequency in $d$ of the format tags (bold, header, italics, underlined, etc) attached with $x$. We skip details of how exactly we do the combination due to lack of space.

Finally, the extracted tables are indexed using Lucene and stored on disk. Each table is treated as a Lucene document with three text fields: header, context, and content. While indexing, we associate boost values of 2, 1.5, and 1 respectively with each of the three fields to control the relative importance of matches in the different fields for the column description queries.

## 2.2 Query Time Processing

Every user query $Q$ consisting of say, $q$ sets of keywords $Q_1, \ldots, Q_q$ is processed through the following stages.

### 2.2.1 Find Candidate Web Tables

We probe the index using the union of words in $Q_1, \ldots, Q_q$. This gives us a set of tables $\mathcal{T}^1 = T_1, \ldots T_{n_1}$ that match the query keywords in the header, context, or body. However, since many Web tables do not contain headers or useful context words, we use a second index probe to retrieve tables $\mathcal{T}^2$ based on content overlap with tables in $\mathcal{T}^1$. But, typically many tables in $\mathcal{T}^1$ are irrelevant, and arbitrarily retrieving all tables with overlapping content in $\mathcal{T}^1$ can introduce too much noise. We deploy the following conservative strategy. Invoke the column mapper (described in the next section) to find the top-two tables with very high relevance score from $\mathcal{T}^1$. We may find no such tables for a query. We then add a random set $S$ of ten rows from the confident tables and make a second index probe with the union of $Q$ and $S$. In our experiments we found that the second stage index probe was used in 65% of the queries in our workload described in Section 5. For these, on an average 50% of all relevant source tables were obtained from the second stage. The fraction of relevant tables in the first stage was just 52% as compared to 70% in the second stage. This shows the usefulness of the second stage index probe.

### 2.2.2 Column Mapping

This step works on the candidate web tables from the index probe $T_1, \ldots T_n$ and decides for each table $T_i$ if it is relevant and if so marks for each query column $Q_\ell$ the column (if any) of $T_i$ that maps to $Q_\ell$. Also, with each relevance and column labeling decision it associates well calibrated probability scores since these are needed to fetch additional tables for the second index probe, and for ranking rows in the final table. The column mapper exploits a number of clues including matches in the header and context of a web table, the overlap in the content of two web tables, and the association of query keywords to a column's corpus in the entire corpus. In Section 3 we present how we combine these clues to solve the column mapping task.

### 2.2.3 Consolidator and Ranker

Given a set of tables output by the column mapper along with table-level relevance scores and column-level confidence scores, the consolidator merges relevant columns and rows into a single table. The main challenge in this task is deciding if two rows from different tables are duplicates, and we refer the reader to [9] for the approach we used to solve this. The ranker reorders the rows of the consolidated table so as to bring more relevant and highly supported rows on top. We skip details of our ranker since the focus of this paper is the column mapping task.

## 3. DESIGNING THE COLUMN MAPPER

We are given a query header $Q$ with $q$ sets of keywords $Q_1, \ldots, Q_q$ and a set of noisy web tables $T_1, \ldots T_n$. Each candidate web table has a variable number of columns and rows, may have zero or more headers, and contextual text extracted from around the table. Our goal in this step is to establish if a web table $t$ is relevant to $Q$, and if so, label each column $c$ of $t$ with the query column to which it maps, or a special label na indicating no match to any query column. We call this the Column mapping task, with the understanding that it also subsumes the task of establishing if a table is relevant.

A simple method of solving this problem is as follows: First, establish the overall relevance of a table $t$ to $Q$ by suitably thresholding the sum of TF-IDF similarity of the keywords in $Q$ to the context and header text of $t$. If $t$ is found relevant, find the best matching of the query columns to the columns of $t$ based on a thresholded similarity of each $Q_\ell$ to the header text of each column of $t$.

This basic method has several shortcomings. First, because the context text associated with a table is a very noisy descriptor of a table, the table-level relevance decision could get misled by unrelated verbosity in the context. Ideally, the table level relevance decisions should be made in conjunction with finding useful column mappings in the table. Second, since the headers in web tables are noisy and/or ambiguous, similarity scores with the header text alone is likely to be unreliable. We will show how to exploit other kinds of clues such as the context and body of the table, and other corpus wide co-occurrence statistics to augment the header-level information. Finally, this method cannot handle tables with no headers, or tables with very little context or header text. In such a case we would like to exploit the content overlap of a column with columns of other tables. We will show in the experimental section that ad hoc ways of including header text from other overlapping table columns fails to yield good results.

Instead of separately deciding for each web table whether it is relevant or not, and to which query label it maps using independent similarity measures, we propose a more global approach to combine the diverse clues both within and across tables. We propose a unified approach based on graphical models that provides an elegant formulation to the task of jointly labeling all columns. We first give a brief overview of graphical models and then present our formulation.

*Graphical Models.* A graphical model [10] expresses the joint distribution over a set of $n$ random variables $\mathbf{x} = x_1, \ldots, x_n$, where each $x_i$ belongs to a space of labels $\mathcal{Y}_i$. The model captures the dependencies between elements of $\mathbf{x}$ as a graph $G$, each a node in the graph, with a sparse set of edges as follows:

We first identify small subsets of variables, called *cliques*, that are highly dependent on each other and form complete subgraphs of $G$. For each clique $C$, we then define potential functions $\theta(C, \mathbf{x}_C) \mapsto$ that provides an un-normalized measure of compatibility among the labels $\mathbf{x}_C$ assigned to variable subset $C$. Two special kinds of potentials are *node potentials* $\theta(i, x_i)$ defined on the label $x_i$ of a single node $i$ and *edge potentials* $\theta(i, x_i, j, x_j)$ defined over edge $(i, j)$ in $G$ and labels $(x_i, x_j)$ assigned to the two nodes it connects. The probability distribution is defined as

$$\Pr(x_1, \ldots, x_n) = \frac{1}{Z} \exp(\sum_{C \subseteq \text{cliques}(G)} \theta(C, \mathbf{x}_C))$$

where $Z$ serves to normalize the product of clique potentials to a proper probability. A common method to define potentials is as a dot product between a *model* parameter and a *feature* vector. In case of node potential, we might write $\theta(i, x_i) = \mathbf{w}_1^\top \mathbf{f}(i, x_i)$, where $\mathbf{f} : [1, M] \to^k$ is the feature vector and $\mathbf{w}_1 \in^k$ is the model vector where $k$ is the number of features. Similarly, an edge potential would be defined as $\theta(i, x_i, j, x_j) = \mathbf{w}_2^\top \mathbf{f}(i, x_i, j, x_j)$. The feature vector are designed by the user whereas the model vectors $\mathbf{w}_1, \mathbf{w}_2$ are *trained* from labeled data. This feature-based framework provides a very convenient way to use a variety of intuitive clues without worrying about how to combine them numerically into a single objective. Once $\Pr(\mathbf{x})$ is defined and the model vectors are trained, the *inference problem* is to find $\text{argmax}_\mathbf{x} \sum_C (\theta(C, \mathbf{x}_C))$, the most likely joint assignment of labels to variables. When the graph has many large cliques, this problem is intractable and it is necessarily to design good approximation algorithms.

We now show how we model the column mapping task as a graphical model by defining the random variables in Section 3.1, the node potentials in Section 3.2, the edge potentials in Section 3.3, other higher order potentials in Section 3.4, and the inference algorithm in Section 4.

### 3.1 Variables

Our task is to label each table as relevant or not, and label the columns of relevant tables with one of $q$ query columns or na. Accordingly, a natural choice would be to create a graphical model where we have a variable $t$ for each web table $t \in \{T_1, \ldots, T_n\}$ and a variable $(tc)$ for each column $c$ of table $t$. Each table variable is binary indicating if the table is marked relevant or not, and each column variable $(tc)$ takes a label from the set $\{1, \ldots, q\} \cup \{\text{na}\}$. However, this natural variable representation gives rise to an unnaturally complicated edge set to capture the interaction between a table variable, and *all* its columns — only when a table variable is labeled relevant, does it make sense to label the column variables. Also, to represent positive affinity among similar columns of only relevant tables we would need cumbersome four-variable potentials. Instead, we chose a representation with only column variables (that we index as $tc$) but augmented the label space of column variables with a label nr to indicate that the column is part of an irrelevant table. Thus, a column variable can take one of $q + 2$ labels from the set $\mathcal{Y} = \{1, \ldots, q\} \cup \{\text{na}, \text{nr}\}$.

### 3.2 Node Potentials

Node potentials $\theta(tc, \ell)$ have to be defined for each column variable $tc$ and each label $\ell \in \mathcal{Y}$. Their representation as a weighted sum of features provides a flexible mechanism to incorporate a wide variety of clues. We define five features that captures a diverse set of clues including the similarity between the query keywords and header, context, and body of web tables, and affinity between query keywords and contents of tables. These are defined in Sections 3.2.1 to 3.2.4.

#### 3.2.1 Matching Query Keywords to Table Columns

The most important signal in mapping a web table column $(tc)$ to a query column $\ell$ is the similarity between the query keywords $Q_\ell$ and the text in table $t$'s context (if any) and text in $c$'s header which could be absent or span over one or more rows. A standard method in IR to assign scores to multi-field documents is to compute the similarity of each field separately to the query keywords and take their weighted sum. Adapting to our case, we could separately measure the similarity of $Q_\ell$ to (1) the header tokens in $(tc)$ and (2) the context of $t$. We next illustrate the various limitations of this method.

First, the header text might contain only a part of the query keywords and the other part might appear in the context. For example, one query in our workload was "Nobel prize winner". This retrieved several tables where the word

"Nobel prize" appears in the context and only "winner" appears in the header. If we separately measure the similarity of $Q_\ell$ to $(tc)$'s header and $t$'s context, we will get low similarity from the header, and even if the similarity with context is high it does not help us decide which of the columns specifically matches $Q_\ell$.

Second, matches with tables having multiple header rows is not well defined. Should the header tokens be concatenated and then matched to $Q_\ell$, or should we compute similarity only with a single best matching row? Concatenation makes sense when a true header is split across multiple rows as in column 3 of Table 1 in Figure 1. However, concatenation is bad when some header rows are wrong, or when a second header row presents irrelevant details. For example, in table 2 of Figure 1, the words "chronological order" in the second header row should not be used to diminish the similarity of the first column of Table 2 to $Q_3$. In this case, going with the single-best option is better.

Third, in some cases the evidence might come from a table's content. For example, a query string $Q_\ell$ in our dataset about "Black metal bands" matched a three column web table with respective headers "Band name | Country | Genre" and no context. The table contains names of many band types, of which "Black metal" is one. The only way to find this table as relevant is to exploit the frequent occurrence of the words "Black metal" in the third column in mapping the first column to $Q_\ell$.

Finally, in some cases the header text of some other column $c'$ of $t$ might be needed to match column $c$ to $Q_\ell$. For example, another query in our dataset was about "dog breeds" which matched a web table with several columns two of which had headers "dog" and "breed". The "dog" column contained dog names, no where else was the token "dog" present in the table.

We present a new similarity measure that removes all the above limitations in one unified function. Instead of separately matching whole of $Q_\ell$ to each field, we compute similarity via a two-part segmentation of $Q_\ell$: one part is used to pin $Q_\ell$ to a specific header of $t$ and the other part is used to gather relevance support from other parts of the table. Let $q_1, \ldots, q_m$ be the sequence of tokens in $Q_\ell$. Let $h$ denote the number of header rows in $t$ and let $H_{rc}$ denote the tokens in header row $r$ of column $c$. We measure the similarity of $Q_\ell$ to a column $c$ of web table $t$ by segmenting $Q_\ell$ into two parts: a prefix $P = q_1, \ldots, q_k$ and a suffix $S = q_{k+1}, \ldots, q_m$. One of the parts, prefix or suffix, is matched to a header row $r$ of $c$, and the other part to portions of $t$ outside the header, including the title of $t$, the context of $t$, frequent content in some column of $t$, other headers in column $c$ ($H_{r'c}$ for $r' \neq r$), and other column headers of $t$ in row $r$ (i.e. $H_{rc'}$ for $c' \neq c$). Let $\text{inSim}(P, H_{rc})$ denote the first similarity, that is, the score of matching the prefix $P$ (or suffix $S$) to the header $H_{rc}$ and $\text{outSim}(S, t, r, c)$ denote the second score, that is, the score of matching the remaining part (suffix $S$ or prefix $P$) with the rest of the table. The segmented similarity $\text{SegSim}(Q_\ell, tc)$ is computed as the maximum weighted sum of scores over all possible values of $r$ and all possible query segmentations. That is,

$$\text{SegSim}(Q_\ell, tc) = \max_{r=1\ldots h} \max_{\substack{P,S \text{ s.t.} \\ P \cap H_{rc} \neq \phi \\ PS=Q_\ell \vee SP=Q_\ell}} \frac{\|P\|^2}{\|Q_\ell\|^2} \text{inSim}(P, H_{rc})$$
$$+ \frac{\|S\|^2}{\|Q_\ell\|^2} \text{outSim}(S, t, r, c) \quad (1)$$

where $\|P\|$ denotes the L2-norm of the TF-IDF vector over the tokens in $P$. The $\text{inSim}(P, H_{rc})$ similarity is the TF-IDF weighted cosine similarity between the token sequence $P$ and $H_{rc}$. The outSim similarity is more challenging because we need to account for five sources of matches in the table: $t$'s title and context, other header rows in column $c$, other header columns in $r$, and frequent body content tokens. For ease of notation we call these parts as $T, C, H_c, H_r$, and $B$ respectively. Matches in these different parts have different degrees of reliability. We characterize this reliability with a probability parameter $p_i$ for $i \in \{T, C, H_c, H_r, B\}$. We calculated these empirically on our workload in Section 5 as follows: for each part $i \in \{T, C, H_c, H_r, B\}$ of all $Q_\ell$ and relevant $t$, reliability $p_i$ of part $i$ is the fraction of correctly matched columns from all columns $c$ with positive inSim and positive match with $i$. These values turned out to be $(1.0, 0.9, 0.5, 1.0, 0.8)$ respectively for $\{T, C, H_c, H_r, B\}$. The score of a token is the soft-max of the reliability over the parts with which it matches. Finally, the $\text{outSim}(S, t, r, c)$ value is computed as the sum of the soft-maxed match reliability of each term $w$ weighted by the TF-IDF score $\text{TI}(w)$ of the term as follows:

$$\text{outSim}(S, t, r, c) = \sum_{w \in S} \frac{\text{TI}(w)^2}{\|S\|^2} (1 - \prod_{\substack{i \in \{T, C, H_r, H_c, B\} \\ w \in \text{part}(i)}} (1 - p_i))$$

The SegSim similarity has several nice properties. First, for multi-row headers it overcomes the shortcomings of the two extremes of full concatenation and single best. When the header has no spurious tokens outside the $r^{th}$ row, it reduces to the desired option of cosine similarity on fully concatenated headers. For the other extreme when all but one header is relevant, it reduces to the single best option. Second, when a token matches a table in multiple parts, the outSim score of the table increases but the influence of each additional match decays exponentially. Third, by requiring that the column header have a non-zero match with either the prefix or suffix of $Q_\ell$, we ensure that table-level matches count only for specific columns rather than all columns of a table.

### 3.2.2 Query Fraction Matched

Another useful feature for defining node potentials is the fraction of query tokens that match the column header and other parts of the table. For the same value of SegSim similarity, we should assign higher score to the case where we have covered all the query terms, over the case where we have covered only a subset. We add a second feature called $\text{Cover}(Q_\ell, tc)$ that differs from SegSim in Equation 1 in only the definition of $\text{inSim}(P, H_{rc})$ — instead of cosine similarity we measure the weighted fraction of $P$'s tokens that appear in $H_{rc}$ ( $= \frac{1}{\|P\|^2} \sum_{w \in P \cap H_{rc}} \text{TI}(w)^2$).

### 3.2.3 Affinity of Query Keywords to Column Content

Instead of depending solely on the Web tables retrieved for a query, we could define features over the entire corpus $D$. One potential signal is the co-occurrence of keywords in $Q_\ell$ with the contents of a table column. Such signals have been used in [2] for ranking tables for topic queries, and in [12] for collecting class instance lists on web queries, and in other NLP tasks [20]. Adapting it to our case of query keywords $Q_\ell$ and contents of column $(tc)$ we define a PMI[2]

score as in [2] as

$$\text{PMI}^2(Q_\ell, tc) = \frac{1}{\#\text{Rows}(t)} \sum_{r \in \text{Rows}(t)} \frac{|\text{H}(Q_\ell) \cap \text{B}(\text{Cell}(t,r,c))|^2}{|\text{H}(Q_\ell)| \, |\text{B}(\text{Cell}(t,r,c))|}$$

where $H(Q_\ell)$ denotes the tables in $D$ that contain keywords $Q_\ell$ in their header or context, and $B(\text{Cell}(t,r,c))$ denotes the tables that match the words in cell $(r, c)$ of table $t$ in their content.

### 3.2.4 Table Relevance

Our final feature for node potentials captures the overall relevance of a table to the union of query keywords. We define this as follows:

$$R(Q, t) = \frac{1}{q} \text{clip}(\sum_{\ell \in 1...q} \max_c \text{Cover}(Q_\ell, tc), \min(q, 1.5)) \quad (2)$$

where the function $\text{clip}(a, b)$ is 0 when $a < b$ and $a$ otherwise. Intuitively, this measures the total fraction of query words that match a table's header and context provided the match fraction is greater than 1 for single column queries, and 1.5 for all other queries.

*Node Potential.* Our final node potential $\theta(tc, \ell) =$

$$\begin{cases} w_1 \text{SegSim}(Q_\ell, tc) + w_2 \text{Cover}(Q_\ell, tc) \\ \quad + w_3 \text{PMI}^2(Q_\ell, tc) + w_5 & \text{if } 1 \leq \ell \leq q \\ w_4 \frac{\min(q, n_t)}{n_t}(1 - R(Q, t)) & \text{if } \ell = \text{nr} \\ 0 & \text{if } \ell = \text{na}. \end{cases} \quad (3)$$

where $n_t$ is the number of columns in table $t$. The values $w_1, \ldots, w_4$ capture the relative importance of different features and $w_5$ is a bias term which is negative and serves the purpose of disallowing maps to a query column based on very small similarity values.

## 3.3 Edge Potentials

Our edge potentials $\theta(tc, \ell, t'c', \ell')$ are defined over pairs of columns $tc$ and $t'c'$ and are used to capture the similarity in the content of columns across tables. A popular choice for edge potentials is the positive POTTS potential that assigns a positive score when the two variables take the same label, and is zero otherwise. The strength of the edge is tied to the similarity between the columns and is defined as:

$\theta(tc, \ell, t'c', \ell') = w_e \text{sim}(tc, t'c')[\![\ell = \ell']\!] \quad \forall \, tc, t'c' \text{ s.t. } t \neq t'$

where $[\![C]\!]$ is 1 when $C$ is true and 0 otherwise, and $w_e$ is the weight of this edge feature. This potential fared poorly because a large fraction of web tables are irrelevant, yet some of their columns are similar to columns of relevant tables. These relevant columns get pulled toward the irrelevant label because of the reward accrued from the many edges with irrelevant columns. We fix this problem by setting $\theta(tc, \ell, t'c', \ell')$ to zero when $\ell = \ell' = \text{nr}$. However, this fix causes the reverse problem: irrelevant tables start getting marked as relevant. When irrelevant tables are large in number and highly similar to each other, the edge potential terms overshadow the node potentials. We therefore designed a custom edge potential that differs from the above potential in three ways:

*Normalize Similarity.* We first normalize the similarity measure so that the sum of the similarity of a column to other neighbors is bounded at one. For each column (tc), we normalize its similarity as $\text{nsim}(tc, t'c') = \frac{\text{sim}(tc, t'c')}{\lambda + \sum_{\bar{t}, \bar{c}} \text{sim}(tc, \bar{t}, \bar{c})}$. The smoothing constant $\lambda$ makes sure that the normalization does not inadequately boost similarity of a column that is weakly similar to only a few other columns. We choose $\lambda = 0.3$ and ignore neighbors with unnormalized similarity below 0.1. The nsim values are asymmetric and we will see shortly how these are used to create symmetric edge potentials.

*Include Column Confidence.* Second, we only add an edge potential across tables where at least one of the two columns is confident about its labeling independent of other tables. We measure confidence of a column's labeling using its node potentials to get a probability distribution $\Pr(\ell|tc)$ as described in Section 4.2. A column is confident only if $\Pr(\ell|tc)$ is large for some $\ell \in [1 \ldots q]$. We used a threshold of 0.6.

*Max-matching Edges.* Third, for a table pair $t, t'$, we connect each column $c$ in the pair to at most one column in the other table instead of all the columns with which $c$ might be similar. This provides more robust transfer of labels when columns within a table are similar to each other. We achieve this by first finding the best one-one matching between the columns of two tables based on a weighted sum of their content and header similarity. We only add an edge between columns in the matching, and not between all possible pairs of similar columns of two tables.

The final edge potential $\theta(tc, \ell, t'c', \ell') =$

$$w_e [\![\ell = \ell' \wedge \ell \neq \text{nr}]\!](\text{nsim}(tc, t'c')[\![\max_{y \neq \text{nr}} \Pr(y|t'c') > 0.6]\!]$$
$$+ \text{nsim}(t'c', tc)[\![\max_{y \neq \text{nr}} \Pr(y|tc) > 0.6]\!]) \quad (4)$$

## 3.4 Table-level Potentials

We define a set of hard constraints to control the consistency of the labels assigned to a web table. Each of these are defined over the labels of all the columns within a table and disallow inconsistent labelings by taking large negative values for inconsistent labels. Unlike for the node and edge potentials, there are no parameters to train for these potentials. We include the following four hard constraints.

*The* MUTEX *Constraint.* This constraint requires that only one column in a table $t$ be mapped to a query column. That is,

$$\phi_a(\ell_1, \ldots, \ell_{n_t}) = [\![\sum_j [\![\ell_j = \ell]\!] \leq 1, \forall \ell = 1 \ldots q]\!]_{-\infty}^0 \quad (5)$$

where $[\![C]\!]_{-\infty}^0$ takes a value zero when condition $C$ is true and $-\infty$ otherwise.

*The* ALL-IRR *Constraint.* This constraint ensures that if one column in a table $t$ is assigned a label nr then all columns of that table must be assigned nr. This constraint helps to make consistent table-level relevance decisions from column-level decisions. That is,

$$\phi_b(\ell_1, \ldots, \ell_{n_t}) = [\![\sum_j [\![\ell_j = \text{nr}]\!] \in \{0, n_t\}]\!]_{-\infty}^0 \quad (6)$$

*The* MUST-MATCH *Constraint.* This constraint requires that every relevant table must contain the first query column.

$$\phi_c(\ell_1, \ldots, \ell_{n_t}) = [\![\sum_j [\![\ell_j \in \{1, \text{nr}\}]\!] > 0]\!]_{-\infty}^0 \quad (7)$$

*The* MIN-MATCH *Constraint.* Every relevant table must contain at least $m$ of the $q$ query columns. We chose $m$ as two for all queries with $q \geq 2$.
$$\phi_d(\ell_1, \ldots, \ell_{n_t}) = [\![\sum_j [\![\ell_j \in [1 \ldots q, \text{nr}]]\!] \geq m]\!]_{-\infty}^0 \quad (8)$$

*Overall Objective.* Thus we have reformulated the Column mapping task as the task of finding labels $y_{tc} \in \mathcal{Y}$ to each table, column pair ($tc$) so as to maximize the sum of node potentials (Equation 3), edge potentials (Equation 4) and table consistency potentials (Equations 5 to 8). We denote the set of all labels as $\mathbf{y}$, and the overall objective is:

$$\max_{\mathbf{y}} \underbrace{\sum_{tc} \theta(tc, y_{tc})}_{\text{node}} + \underbrace{\sum_{tc} \sum_{t' \neq t, c'} \theta(tc, y_{tc}, t'c', y_{t'c'})}_{\text{edge}}$$
$$+ \underbrace{\sum_{C \in \{a,b,c,d\}} \sum_t \phi_C(y_{tc_1}, \ldots, y_{tc_{n_t}})}_{\text{constraints}} \quad (9)$$

The above objective has six parameters $w_1, \ldots, w_5, w_e$. We use a training phase with a separate labeled dataset to find the values of these parameters so that the error of the highest scoring mapping is minimized. Since we had only six parameters, we were able to find the best values through exhaustive enumeration. More sophisticated training methods, say based on max-margin structured learning [10] perform well only under exact inference which is not possible for our objective because it is NP-hard. This can be proved using a reduction of the well-known NP hard problem of metric labeling [6] to our problem.

## 4. INFERENCE ALGORITHMS

Since the problem of finding the optimal column labels as per objective 9 is NP-hard, we present approximation algorithms in this section. We develop our approximation algorithm in stages. First in Section 4.1, we show that if the edge potential terms are absent, we can find the optimal column labels in polynomial time via a novel reduction to a bipartite matching problem. Next, we show two approaches to approximating the overall objective: table centric (Section 4.2) and edge-centric (Section 4.3) based on whether they give more importance to the table-specific potentials or across table edge potentials.

### 4.1 Table Independent Inference

We show how to optimally solve objective 9 without the edge potentials. The optimal labeling of the different tables now get decoupled. For each table $t$ we solve a generalized maximum matching problem in a bipartite graph $G_t$ created as follows. The $n_t$ columns of $t$ forming the left nodes of $G_t$ and labels $1, \ldots, q$, na forming the right nodes. The weight of an edge between a left node $c$ and a right node $\ell$ is $\theta(tc, \ell) + M_\ell$ where $M_\ell$ is a large positive constant for $\ell = 1$, and zero otherwise. In addition, each node $v$ is assigned a capacity $c_v$ that is one for all nodes except the right node na for which is $c_v = \max(0, n_t - m)$ where $m$ is the minimum number of columns required to be matched in the MIN-MATCH constraint. Find a matching, that is a subset $S$ of the edges in $G_t$, such that each node $v$ has no more than $c_v$ adjacent edges in $S$ and the sum of edge weights in $S$ is maximized. Such a matching can be found in $O(n_t^2 q)$ time using the well-known reduction to the min-cost max-flow problem (summarized in Section 4.2.1). The edges in the optimal $S$ give the highest scoring column mapping $\mathbf{y}_t$ of $t$ under the constraint that $t$ is relevant. We contrast the score of $\mathbf{y}_t$ with the score of labeling all columns of $t$ as nr and pick the higher of the two as the final labeling.

We skip a formal proof of correctness and instead state some intuitive claims as justification. The capacity one constraint on all right side nodes in $1, \ldots, q$ ensures that the MUTEX condition is met. The large additive constant $M_1$ ensures that the highest scoring labeling must assign at least one column to label 1 thereby meeting the MUST-MATCH constraint. The constraint that no more than $n_t - m$ columns of $t$ can be assigned label na ensures that at least $m$ of them must be mapped to a query column, thus satisfying the MIN-MATCH constraint.

### 4.2 Collective Inference: Table-centric

This collective inference algorithm is centered on optimal inference at the table-level, and uses edges to influence table-level decisions very cautiously. The algorithm works in three stages. First, independently for each table $t$ find the highest possible score $\mu_{tc}(\ell)$ of objective 9 (ignoring edge potentials) when each column $c$ is constrained to take each label $\ell$. Normalize these to define a distribution as $p_{tc}(\ell) = \frac{\exp(\mu_{tc}(\ell))}{\sum_{\ell'} \exp(\mu_{tc}(\ell'))}$. Second, for each table column $tc$ collect messages from each of its neighbors as $\text{msg}(tc, \ell) = \sum_{t'c' \in \text{nbr}(tc)} w_e \text{nsim}(tc, t'c') p_{t'c'}(\ell)$. Finally, independently for each $t$ invoke the algorithm in Section 4.1 with node potential modified as $\max(\text{msg}(tc, \ell), \theta(tc, \ell))$.

The only difficult part of the above algorithm is the efficient computation of the $\mu_{tc}(\ell)$ values. The direct method of computing $\mu_{tc}(\ell)$ by repeated calls to the maximum matching algorithm of Section 4.1 for each $(c, \ell)$ pair is expensive. We propose a method of making this efficient by simultaneously computing all the values. Before we present our algorithm we need to briefly recap the popular flow-based algorithm for maximum matching in bipartite graphs. Readers familiar with this topic can skip Sections 4.2.1 and 4.2.2 and jump to Section 4.2.3.

#### 4.2.1 Recap: Maximum Weight Bipartite Matching

Given a weighted bipartite graph $B$ where each node $u$ has a positive capacity $c_u$ and each edge $(u, v)$ has weight $w(u, v)$, find subset $S$ of edges of $B$ such that $\sum_{(u,v) \in S} w(u, v)$ is maximized and $\sum_{v:(u,v) \in S \vee (v,u) \in S} 1 \leq c_u$. A well-known algorithm [13] to solve this problem is by finding the maximum flow on a weighted directed graph $G$ created from $B$ as follows. First, we need to ensure that the total capacity of nodes on the left side of $B$ is equal to the total capacity on the right. If this condition is violated we add a dummy node $d$ on the deficient side with the deficient capacity. Call the new set of left nodes $L$ and right nodes $R$. $G$ is seeded with these nodes and two extra nodes: a source node $s$, and a sink node $t$. Second, add edges from $s$ to each $u \in L$ with $\text{cost}(s, u) = 0$ and $\text{cap}(s, u) = c_u$, from each $u \in R$ to $t$ with $\text{cost}(u, t) = 0$ and $\text{cap}(u, t) = c_u$, and from each $u \in L$ to each $v \in R$ with $\text{cap}(u, v) = \min(c_u, c_v)$ and $\text{cost}(u, v) = 0$ if either $u$ or $v$ is the dummy node $d$, else $\text{cost}(u, v) = -w(u, v)$. Invoke the algorithm described in the next section to find the minimum cost maximum flow on $G$. The final matching consists of all edges from $L$ to $R$ with positive flow.

#### 4.2.2 Recap: The Max-flow Algorithm

The input to the algorithm is a directed graph $G$ with edge costs $\text{cost}(u, v)$ and $\text{cap}(u, v)$ and two special vertices

**Input**: $G_t$ = Weighted bipartite graph (Section 4.2.3)
$G_f^*, m^*$, Opt = Residual graph and optimal matching obtained by algorithm in Section 4.2.1
**for** $\ell \in [1 \ldots q] \cup \{\text{na}\}$ **do**
    Find shortest distances $d(\ell, .)$ in $G_f^*$ from $\ell$ to all nodes.
    $\mu_{tc}(\ell) = \text{Opt} - d(\ell, c) - \text{cost}(c, \ell), \quad \forall c = 1, \ldots, n_t$
**end for**
$\mu_{tc}(\text{nr}) = \sum_c \theta(tc, \text{nr})$

**Figure 3: Finding all max-marginals in $G_t$.**

$s, t$. The goal is to push the maximum flow from $s$ to $t$ such that the flow $f(u, v)$ along each edge is $\leq \text{cap}(u, v)$ and the total cost $\sum_{(u,v)} \text{cost}(u, v) f(u, v)$ is minimized. The Max-flow algorithm [13] starts with associating each edge $(u, v)$ with $f(u, v) = 0$. Define with each edge $(u, v)$ a residual capacity $\text{res}(u, v) = \text{cap}(u, v) - f(u, v)$ that measures how much extra flow can be pushed along that edge without violating the capacity constraint. For every $(u, v)$ in the graph, there is an implicit reverse edge $(v, u)$ with $\text{cap}(v, u) = 0$, $f(v, u) = -f(u, v)$, and $\text{cost}(v, u) = -\text{cost}(u, v)$. This implies that a flow along $(u, v)$ can be reversed by pushing a flow in the reverse direction from $v$ to $u$. Every set of flow values $f$ defines a new residual graph $G_f$ out of $G$ comprising only of edges with positive residual capacity. The algorithm proceeds by repeatedly finding the shortest cost path $P$ from $s$ to $t$ in $G_f$, pushing the maximum possible flow $f(P)$ along $P$ in $G$, thereby modifying $f$ and getting a different residual graph from $G$. The algorithm terminates when no path can be found from $s$ to $t$ in $G_f$.

### 4.2.3 Finding Max-marginals

We now show how to efficiently compute $\mu_{tc}(\ell)$ the maximum possible score of Equation 9 (without the edge potentials) under the constraint that column $c$ takes label $\ell$. These are called max-marginals. When computing the max-marginals, it is important to exclude the MUST-MATCH and MIN-MATCH constraints because otherwise the relative magnitude of $\mu$ for different $\ell$ can be distorted when some high scoring labeling violates the MUST-MATCH or MIN-MATCH constraint. Thus, for each $t$ $\mu_{tc}(\ell) =$

$$\max_{\mathbf{y}: y_{tc} = \ell} \sum_{c'} \theta(tc', y_{tc'}) + \phi_a(y_{tc_1}, \ldots, y_{tc_{n_t}}) + \phi_b(y_{tc_1}, \ldots, y_{tc_{n_t}}) \quad (10)$$

We create a bipartite graph $G_t$ as in Section 4.1 with the only difference that the addition $M_1$ to edge weights is left out since the MUST-MATCH constraint is dropped and the capacity of the right node na is $n_t$ since the MIN-MATCH constraint is dropped. We then find the maximum weight bipartite matching (recalled in Section 4.2.1) on $G_t$. Let $m_c^*$ denote the node to which $c$ is matched in the optimal matching $m^*$, Opt denote the sum of edge weights in the optimum matching, and let $G_f^*$ be the final residual graph from the Max-flow algorithm (recalled in Section 4.2.2). For each $c$, and for each $\ell \neq m_c^*$, we need to force a matching $(c, \ell)$ and compute the rest of the matching. Since the capacities of the two sides were balanced in $m^*$, we need to remove unit flow from $m_\ell^*$ to $\ell$ and from $c$ to $m_c^*$ and add flow from $c$ to $\ell$. The smallest cost method to make this change is by reverting the flow along the *shortest* path $P$ from $\ell$ to $c$ in $G_f^*$. Let $d(\ell, c)$ be the cost of this path. The maximum weight matching under the $(c, \ell)$ constraint can be shown to be equal to $\text{Opt} - d(\ell, c) - \text{cost}(c, \ell)$. Thus, in order to find the maximum matching under all possible pair constraints, we need to find single source shortest paths from each right node $\ell$ in the residual graph. Since edge costs can be negative, we need to use the Bellman Ford shortest path algorithm to find these shortest paths. Each invocation of Bellman Ford's algorithm on a graph with $N$ nodes and $E$ edges takes $O(NE)$ time, and we do this $q + 1$ times. A pseudo code of the algorithm appears in Figure 3.

## 4.3 Collective Inference: Edge-centric

The table-centric algorithm gives more importance to table-level constraints than to edge potentials. We wanted to compare this approach with an opposite approach of giving edge potentials central importance. In these approaches, called edge-centric approaches, the table-level constraints are either expressed as edge potentials, or are handled in a post-processing phase. Another reason for exploring these approaches is that there are many existing inference algorithms that work on graphical models with only node and edge potentials [10, 1, 11, 18]. These include message passing algorithms like belief propagation [10], TRWS [11], and MPLP [18], and graph cut-based algorithms like $\alpha$-expansion algorithm [1]. Of these, the $\alpha$-expansion algorithm is known to be very efficient [19] and was the best performing of the edge-centric algorithms in our experiments. In this section we go over how we adapted the algorithm to handle the table constraints.

The $\alpha$-expansion algorithm only works for edge potentials whose negated forms behave like a metric. That is, for each pairwise edge potential function $\phi$ it requires that $\phi(\ell_i, \ell_j) = \phi(\ell_j, \ell_i)$, and $\phi(\ell_i, \ell_j) - \phi(\ell_j, \ell_k) \leq \phi(\ell_i, \ell_k)$. It is easy to verify that our edge potentials in Equation 4 satisfy this property.

Also, of the four constraints, the ALL-IRR constraint in Equation 6 can be written as the sum of metric edge potentials defined over each pair of columns, that is, $\phi_b(\ell_1, \ldots, \ell_{n_t})$ can be expressed as $\sum_{j=1}^{n_t} \sum_{i=1}^{j-1} \phi_B(\ell_i, \ell_j)$ where

$$\phi_B(\ell_i, \ell_j) = [\![[\![\ell_i = \text{nr}]\!] + [\![\ell_j = \text{nr}]\!] \neq 1]\!]_{-\infty}^0 \quad (11)$$

We first present an overview of the $\alpha$-expansion algorithm and then show how we modify it to incorporate the remaining three of the table constraints.

*The $\alpha$-expansion Algorithm [1].* The algorithm maintains a labeling $\mathbf{y}$ of the variables at all times, and in each step improves it through local moves. Initially $\mathbf{y}$ can be arbitrary — say, all vertices assigned label na. Next, for each label $\alpha$ ($\alpha \in \mathcal{Y}$) an $\alpha$-expansion move switches the labeling of an optimal set of vertices from their current label in $\mathbf{y}$ to $\alpha$. This is repeated in a round over the $q + 2$ labels until $\mathbf{y}$ remains unchanged in a complete round. For graphs whose edge potentials satisfy the metric condition, an optimal $\alpha$-expansion move is obtained from the minimum cut in a weighted directed graph $G_{\alpha, \mathbf{y}}$ created as follows: To the original graphical model $H$ add two special vertices $s, t$. Add an edge from $s$ to each $v \in V(H)$ and from every $v \in V(H)$ to $t$. Edge weights are derived from the potentials of $H$, the current $\alpha$ and current labeling $\mathbf{y}$ such that the minimum $s$-$t$ cut gives the optimal subset of variables $S$ on the $t$ side whose labels in $\mathbf{y}$ should be switched to $\alpha$. We refer the reader to [1] for details on how to set these edge weights.

We next present our modification to the above algorithm to handle the remaining constraints. We will shortly present a novel modification for handling the MUTEX constraint. The MUST-MATCH and MIN-MATCH constraints cannot be easily

**Input**: $G, s, t$ and disjoint vertex groups $V_1, \ldots V_T$.
$G_f, S$ = Residual graph and $t$ side vertices after applying Max-flow to find the minimum $s$-$t$ cut on $G$
**while** $\exists\ k\ \ s.t.\ \ |S \cap V_k| > 1$ **do**
  **for all** $i\ \ s.t.\ \ |S \cap V_i| > 1$ **do**
    **for** $v \in (U_i = S \cap V_i)$ **do**
      $f(v, V_i)$ = maximum additional flow in $G$ if for each $u \in U_i - \{v\}$, $\text{cap}(s, u)$ is increased to $\infty$.
    **end for**
  **end for**
  Pick $i^*, v^* = \text{argmin}_{i,v} f(v, V_i)$, set $\text{cap}(s, u) = \infty$ for each $u \in U_{i^*} - \{v^*\}$ and modify flow in $G_f$
  $S$ = vertices on the $t$ side of the cut
**end while**
**Output**: $S$

**Figure 4: Constrained minimum $s$-$t$ cut algorithm.**

integrated with the above algorithm. So, we handle these as a post-processing step as follows: In the output labeling if any table $t$ violates the MUST-MATCH or MIN-MATCH constraint we greedily fix its labels by invoking the table independent algorithm (Section 4.1) on $t$.

To handle the MUTEX constraint, we modify the $\alpha$-expansion move for $\alpha \in [1 \ldots q]$ so that at most one column variable in each table switches to $\alpha$. We achieve this by solving the following constrained $s$-$t$ cut problem on $G_{\alpha,\mathbf{y}}$ for $\alpha \in [1 \ldots q]$.

*The Constrained Minimum $s$-$t$ cut Problem.* Given a positive weighted directed graph $G = (V, E, \mathbf{w})$ whose vertex set can be partitioned into disjoint subsets $V_1, \ldots V_T$ and two special vertices $s, t$, find the minimum $s$-$t$ cut such that at most one vertex in each $V_i$ is on the $t$ side of the cut. In our case the groups $V_i$ correspond to columns of the same table. This problem is NP-hard unlike the unconstrained minimum $s$-$t$ cut problem which can be solved in $O(VE \log V)$ time. We are able to provide a factor of two approximation. In this paper we skip proofs. Instead we present an approximation that performed well on our problem.

A popular method for solving the minimum $s$-$t$ cut problem is based on a reduction to the Max-flow problem that we reviewed in Section 4.2.2. We first show this reduction and then present our modification.

*Minimum $s$-$t$ cut via Max-flow[13].* The Max-flow algorithm is invoked on $G$ with a uniform cost of one on all edges and $\text{cap}(u, v) = w_{uv}$ where $w_{uv}$ denotes the weight of an edge in $G$. Let $G_f$ be the residual graph when the Max-flow algorithm terminates. The minimum cut $\mathcal{C}$ comprises of edges for which $f(u, v) = \text{cap}(u, v)$.

*Our Modifications for Constrained Cuts.* Let $S$ be the set of vertices that remain connected to $t$ after the edges $\mathcal{C}$ in the unconstrained cut are removed. $\mathcal{C}$ is optimal for the constrained case if for each vertex group $V_i$ at most one member appears in $S$. Else we repeat the following steps until all constraints are satisfied. Find a violated group $V_i$ for which $|S \cap V_i| > 1$. Let $U_i = S \cap V_i$. We force all but one vertex $v^*$ in $U_i$ to the $s$ side of the cut. Such a $v^*$ is picked as that vertex for which we can push the minimum additional flow in the residual graph $G_f$ with all $U_i - \{v^*\}$ connected to $s$ via infinite capacity edges. The final algorithm appears in Figure 4.

## 5. EXPERIMENTS

Creating a useful query set for this project is challenging in an academic setting without access to a user base. The best we found was the list of topic queries in [2] collected via a Amazon Mechanical Turk service. They represent an unbiased workload because these are suggested by a diverse set of web users, and not hand picked by researchers. The query set comprised of 51 queries spanning topics such as, "north american mountains", "professional wrestlers", and "world's tallest buildings". We call this the AMT query set. However, since our task required multi-column queries, we converted each topic query in AMT to a multi column query as follows: For each topic query in AMT, we inspected the top-10 pages from a Google search and found between one and three prominent attributes to be attached with each query. The final list of multi-column queries is present in Table 1 [1]. The original topic queries can be found in Table 1 of [2]. We augmented the AMT queries with twelve more queries that we collected internally from Wikipedia tables. This gave us a total of 59 queries of which 5 were single-column queries, 37 were two column queries, and 17 were three column queries. We list the 59 queries in Table 1.

We run these queries on a web crawl of 500 million pages from which we found 25 million data tables. For evaluating the accuracy of the Column mapping task we posed each query in our query-set on WWT and for each table returned by the index probe we collected the correct labeling by manually labeling each of the 1906 web tables over the 59 queries. This gave us the set of ground truth labels. Each labeling was reviewed by two human labelers to avoid any human errors. In Table 1 we list the total number of source web tables that were returned by the two-phase index probe in WWT and the number of relevant web tables. Overall for a query, we found between 0 and 68 web tables with an average of 32.29. The fraction of relevant matches varied between 0 and 100% and on average only 60% of web tables were relevant.

We evaluate different methods via the well-known F1 error measure. This is calculated as follows: Let $\mathbf{y}^*$ denote the vector of all column labelings in the ground truth. The F1 error of any other labeling $\mathbf{y}$ produced by a method is

$$\text{error}(\mathbf{y}, \mathbf{y}^*) = 1 - \frac{\sum_{tc} 2[\![y_{tc} = y_{tc}^* \wedge y_{tc} \in [1 \ldots q]]\!] \times 100}{\sum_{tc}[\![y_{tc} \in [1 \ldots q]]\!] + \sum_{tc}[\![y_{tc}^* \in [1 \ldots q]]\!]}$$

We compare the accuracy of the Column mapping task on the following methods

1. Basic: A baseline method that is described in the beginning of Section 3.
2. NbrText: Basic augmented to include text of similar columns to measure similarity $\text{sim}(Q_\ell, tc)$ as
$$\max(\text{TI}(Q_\ell, tc), \max_{t'c'} \text{sim}(tc, t'c')\text{TI}(Q_\ell, t'c'))$$
3. PMI[2]: The basic method augmented with only PMI[2] scores defined in Section 3.2.3
4. WWT: Our graphical model-based approach described in Section 3 with the table-centric algorithm.

For reporting results, we partitioned the queries into two sets: an "easy" set for which all methods were within 0.5% of each other. These include queries for which either the column labeling task is trivial, or impossible to solve with

---
[1]We skipped 4 queries from AMT, because we could not interpret the appropriate column queries for them.
916

## Table 1: Query set.

| Single Column Queries | Source Tables Total | Relevant |
|---|---|---|
| dog breed | 68 | 66 |
| kings of africa | 26 | 0 |
| phases of moon | 56 | 17 |
| prime ministers of england | 35 | 3 |
| professional wrestlers | 52 | 52 |
| **Two Column Queries** | **Total** | **Relevant** |
| 2008 beijing Olympic events \| winners | 29 | 0 |
| 2008 olympic gold medal winners \| sports/event | 26 | 0 |
| australian cities \| area | 30 | 4 |
| banks \| interest rates | 51 | 34 |
| black metal bands \| country | 39 | 19 |
| books in United States \| author | 6 | 2 |
| car accidents location \| year | 46 | 8 |
| clothing sizes \| symbols | 20 | 0 |
| composition of the sun \| percentage | 50 | 12 |
| country \| currency | 56 | 53 |
| country \| daily fuel consumption | 38 | 14 |
| country \| gdp | 58 | 56 |
| country \| population | 58 | 55 |
| country \| us dollar exchange rate | 52 | 43 |
| fifa worlds cup winners \| year | 49 | 9 |
| Golden Globe award winners \| year | 23 | 19 |
| Ibanez guitar series \| models | 21 | 3 |
| Internet domains \| entity | 10 | 4 |
| James Bond films \| year | 16 | 11 |
| Microsoft Windows products \| release date | 25 | 12 |
| MLB world series winners \| year | 13 | 3 |
| movies \| gross collection | 57 | 57 |
| name of parrot \| binomial name | 11 | 8 |
| north american mountains \| height | 47 | 28 |
| pain killers \| company | 1 | 1 |
| pga players \| total score | 40 | 29 |
| pre-production electric vehicle \| release date | 3 | 0 |
| running shoes model \| company | 11 | 5 |
| science discoveries \| discoverers | 41 | 37 |
| university \| motto | 7 | 5 |
| us cities \| population | 34 | 32 |
| us pizza store \| annual sales | 35 | 1 |
| usa states \| population | 41 | 37 |
| used cellphones \| price | 29 | 0 |
| video games \| company | 30 | 28 |
| wimbledon champions \| year | 38 | 24 |
| world tallest buildings \| height | 51 | 12 |
| **Three Column Queries** | **Total** | **Relevant** |
| academy award category \| winner \| year | 56 | 22 |
| bittorrent clients \| license \| cost | 0 | 0 |
| chemical element \| atomic number \| atomic weight | 33 | 30 |
| company \| stock ticker \| price | 53 | 53 |
| educational exchange discipline in US \| number of students \| year | 13 | 2 |
| fast cars \| company \| top speed | 34 | 29 |
| food \| fat \| protein | 47 | 43 |
| ipod models \| release date \| price | 44 | 16 |
| name of explorers \| nationality \| areas explored | 19 | 13 |
| NBA Match \| date \| winner | 44 | 34 |
| new Jedi Order novels \| authors \| year | 25 | 24 |
| Nobel prize winners \| field \| year | 12 | 10 |
| Olympus digital SLR Models \| resolution \| price | 11 | 3 |
| president \| library name \| location | 8 | 1 |
| religion \| number of followers \| country of origin | 37 | 32 |
| Star Trek novels \| authors \| release date | 8 | 8 |
| us states \| capitals \| largest cities | 32 | 30 |

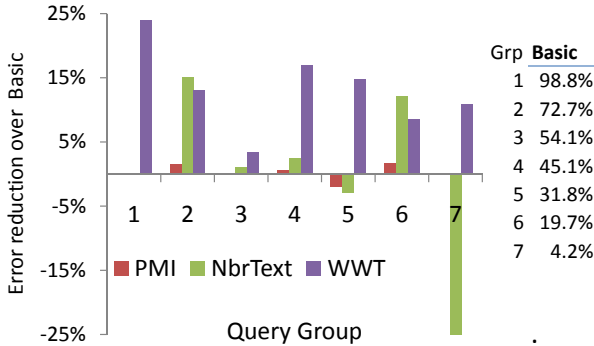

**Figure 5:** Error reduction relative to Basic in the column map performance of WWT, NbrText and $PMI^2$. The table alongside shows the error of Basic.

our set of clues. One-third of our queries were "easy" and all methods had an error close to 22% on these queries. We compare different methods on the remaining two-third "hard" queries. For ease of plotting, we divided the hard queries into seven groups by binning on the error of the Basic method.

We first compare the above four methods. We then analyze the impact of the segmented similarity measure and finally compare different collective inference algorithms.

### 5.1 Overall Comparison

In Figure 5 we show the reduction in error relative to Basic of the three methods: $PMI^2$, NbrText, and WWT on the seven query groups. Overall, WWT incurred an error of 30.3% in contrast to 34.7% for Basic and $PMI^2$, and 34.2% for NbrText. For the first query group (comprising of seven queries), all three methods failed to label all but one as relevant whereas WWT got 22% of them. Even for the last group where Basic already achieved a low error of 4.2, WWT reduced that by 10% to 3.6%. The NbrText method reduces error for some queries but incurs an increase in error in several others. The method is specifically bad when the columns within a table overlap. An example is the last query in Table 1. The NbrText method suffers because the columns containing capitals and largest cities are overlapping, causing the wrong header text to be imported in a column.

Surprisingly, we did not get any accuracy boost overall with the $PMI^2$ score unlike what is reported in [2]. Even if we restrict to single column queries in our set, WWT's error is 28.9% compared to 33.3% of $PMI^2$ used in [2]. One reason is that the PMI scores are noisy — while they reduce error in seven of the queries, they also caused an increase in an equal number mostly due to adding irrelevant columns. Other studies [20] have attributed such behaviour to the undue importance the PMI score gives to low frequency words due to their presence in the denominator. Also, the $PMI^2$ score is expensive to compute. The average time for a query was 6.3 seconds for Basic, 40 seconds for $PMI^2$ and 6.7 seconds for WWT, which does not use the $PMI^2$ scores by default.

We show the impact of improved column mapping on overall search performance in Figure 6 where the y-axis is the error comparing the rows of the consolidated answer table of true column mapping and the consolidated answer table of any other mapping. We observe that WWT yields significant improvements in the accuracy of the final answer in all cases.

In Figure 7 we show the total running time of WWT broken into the time taken to probe the index (stored on disk) in each of the two stages, the time to read and parse the raw tables from disk, the time for column mapping, and the time to consolidate (and dedup) the rows of relevant tables. The running time varies from 1.5 to 14 seconds with an average of 6.7 seconds. The key factors affecting running time are whether a second index probe was used for the query, the size of the raw tables and the number of relevant rows that are consolidated into the final answer. The time for column mapping is a negligible fraction of the total time.

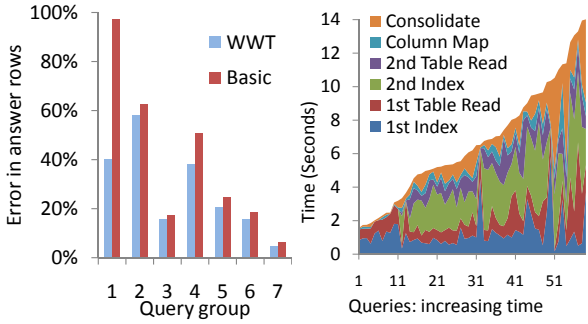

Figure 6: Answer quality.

Figure 7: Running time.

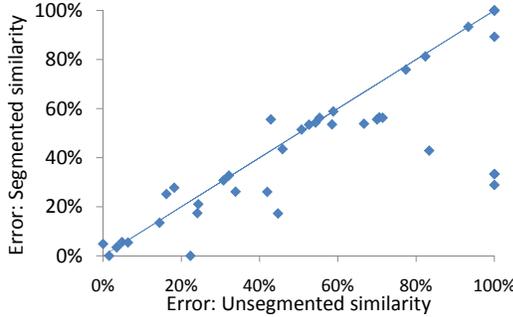

Figure 8: Comparing segmented similarity with standard unsegmented IR similarity.

## 5.2 Evaluating Segmented Similarity

In this section we evaluate our segmented similarity measure of Section 3.2.1 by comparing it against a model which is identical in all respects except using plain cosine similarity with the header text instead of SegSim and Cover. We call this the unsegmented similarity measure. For each measure we retrained the model parameters so that they are best tuned for the selected similarity measure. Overall the segmented measure reduced error from 33.3% to 30.3%. In Figure 8 we show a scatter plot where each point represents the error of a query using segmented measure against error of the unsegmented measure. We see that for all but three of the 32 queries, the segmented approach lies below the 45 degree line, and in eight cases it reduces by more than 10%.

## 5.3 Comparing Collective Inference Methods

In Table 2 we compare a baseline (in column 2) that independently labels each table with the following collective inference algorithms: the constrained $\alpha$-expansion algorithm, two edge-centric message-passing algorithms viz., Belief propagation (BP) and TRWS, and the Table-centric algorithm. The BP and TRWS incorporated the table constraints in the same way as $\alpha$-expansion for all but the MU-TEX constraint. The MUTEX constraints were reduced to edge potentials like the ALL-IRR constraint — we skip details due to lack of space.

| Group | None | $\alpha$-exp | BP | TRWS | Table-centric |
|---|---|---|---|---|---|
| 1 | 76.0 | 76.0 | 74.5 | 76.0 | 75.1 |
| 2 | 72.4 | 72.4 | 71.6 | 72.4 | 63.3 |
| 3 | 53.3 | 43.4 | 61.1 | 51.4 | 52.3 |
| 4 | 43.2 | 45.4 | 40.4 | 41.5 | 37.5 |
| 5 | 31.2 | 26.6 | 26.9 | 31.3 | 27.1 |
| 6 | 22.4 | 20.1 | 20.1 | 18.8 | 18.0 |
| 7 | 4.0 | 4.0 | 4.2 | 4.0 | 3.8 |
| Overall | 33.1 | 31.3 | 31.5 | 32.3 | 30.3 |

Table 2: Comparing different collective inference algorithms on F1 error over seven query groups separately and over all queries combined.

Overall, the Table-centric algorithm achieves the lowest error of 30.3% which is roughly 3% lower than the baseline of no collective inference. This is followed by our constrained $\alpha$-expansion algorithm. BP is slightly worse and TRWS is the worst. One reason is that BP and TRWS are known not to provide good approximations when many edge potentials are dissociative (i.e., they prefer connected nodes to take different labels). In our graphical model, in expressing the MUTEX constraints as edge potentials we created many dissociative edges. In contrast, for the $\alpha$-expansion algorithm we handled the MUTEX constraints separately via a constrained graph-cut algorithm. In order to understand the reasons for the poorer performance of $\alpha$-expansion vis-a-vis the table-centric algorithm, we considered two explanations. First, that the edge potential scores in our objective were not well calibrated causing higher scoring labelings to not necessarily have lower error, and the Table-centric was winning by giving lower importance to such potentials. Second, that the $\alpha$-expansion algorithm was solving the objective (with all the constraints) poorly. We found the latter to be the reason. In most cases where $\alpha$-expansion lost it returned labelings with lower overall scores than the table-centric algorithm. In terms of running time, the table-centric algorithm is the fastest followed by $\alpha$-expansion which is a factor of five slower, followed by BP a factor of six slower and finally TRWS which is a factor of 30 slower. Thus, the most practical option for collective inference is the table-centric algorithm both in terms of accuracy and running time.

## 6. RELATED WORK

We discuss four areas of related work: structured Web search, Web tables in structured search, keyword queries on databases, and schema matching.

Structured Web search has been a topic of research ever since the advent of the Web (see [5] for a survey). The most common type of structured search is *point* answers to a query such as: "CEO of IBM"; and most research has focused on harnessing document sources to answer them [5]. Our focus is on queries whose answer is a *single* consolidated table. In earlier work [9] we developed a query-by-example paradigm of extracting such tables from lists on the Web (web tables were excluded). Another example is Google squared[2], a commercial product, that interpreted query keywords as description of entity types and the answer was a table with entity instances and a suggested list of attributes. Technical details of the system are not in the public domain.

The potential of web tables as a source of structured information was first highlighted in [4, 3]. In this work, the stress was on collecting offline information of various kinds, including attribute synonyms, attribute associations, etc. In contrast, our goal is to answer ad hoc queries, for which the closest related work is [2]. Their system is based on multiple user interactions: first in response to a single set of keywords a ranked list of tables is retrieved, the user integrates the sources into a single table, the user can then choose additional attribute columns, and the system fills in the values from additional sources. Only the method used for establishing table relevance in the first step above is related to our goal in this paper. The method they propose for relevance ranking is to use the PMI score that we have already shown in Section 5.1 to not be as effective for our task of column labeling.

---

[2] http://en.wikipedia.org/wiki/Google_Squared

Keyword search on tables is now an established area of research [21]. Most early work was on clean databases where each entity type has a distinct table with well defined column names, types, and primary keys. Recently [17] presents a probabilistic algorithm for annotating parts of keyword queries with table names, attribute names, and selection predicates on a set of product catalogs. A related problem is tackled in [15] where a keyword query over an Ontology is broken into a structured query over the entity, types, and relationships in the Ontology. A database of web tables is entirely different from such databases: there is huge redundancy, no well-defined schema, no standard syntax for specifying column names, and the scale is orders of magnitude higher.

Our task could be viewed as a schema matching problem between the query columns $Q$ and a Web table $T$. Schema matching [16, 8] has traditionally been applied for integrating two databases, each of which contains a consistent and clean set of tables and the main challenge is in managing the complex alignment between the large number of schema elements on each side. In contrast, in our case we are matching a few query columns to a large number of unlinked and noisy web tables. This gives rise to a very different problem structure —- since a small fixed number of query columns are matched we can cast this matching task as a column labeling task. The building blocks used for matching a query table to any single web tables is similar to those used in schema matching but collectively performing such binary matchings over several web tables is new to our setting.

## 7. CONCLUSION

We presented the design of a system for getting as answer a multi-column table in response to a query specifying a set of column keywords. We described the many non-trivial steps required in the processing of such queries such as extracting headers and context of tables, retrieving tables via a two staged index probe, and mapping columns before consolidating them into the answer table. The focus of this paper was the column mapping task. Our representation of this task as a graphical model allowed us to jointly decide on the relevance and mappings of all candidate table columns while exploiting a diverse set of clues spanning matches in different parts of the table, corpus-wide co-occurrence statistics, and content overlap across table columns. We presented a novel method of matching query keywords to table columns via a two part query segmentation method, and a robust mechanism of exploiting content overlap across table columns. Experiments on a realistic query workload and a database of 25 million web tables showed a 12% reduction in error relative to a baseline method. Future work in the area include exploiting newer corpus wide co-occurrence statistics, alternative structured sources such as ontologies, and enhancing the search experience via faceted search and user feedback.